\documentclass[pdflatex,sn-basic]{sn-jnl}

\usepackage{graphicx}%
\usepackage{multirow}%
\usepackage{amsmath,amssymb,amsfonts}%
\usepackage{amsthm}%
\usepackage{mathrsfs}%
\usepackage[title]{appendix}%
\usepackage{xcolor}%
\usepackage{textcomp}%
\usepackage{manyfoot}%
\usepackage{booktabs}%
\usepackage{algorithm}%
\usepackage{algorithmicx}%
\usepackage{algpseudocode}%
\usepackage{listings}%

\theoremstyle{thmstyleone}%
\theoremstyle{thmstyletwo}%

\theoremstyle{thmstylethree}%

\begin{document}

\title[Article Title]{The B-Index as a Diagnostic of Cool Stars: Assessing Metallicity Dependence of TiO Absorption in the Visible Spectrum}


\author*[1]{\fnm{Fatemeh} \sur{Azizi}}\email{f.azizi@pnu.ac.ir}

\author[2]{\fnm{Mohammad Taghi} \sur{Mirtorabi}}\email{torabi@alzahra.ac.ir}

\author[1]{\fnm{Zahra} \sur{Mohammadian}}\email{}

\affil[1]{\orgdiv{Department of Physics},\orgname{ Payame Noor University},\orgaddress{\state{Tehran},\country{Iran}}}

\affil[2]{\orgdiv{Department of Fundamental Physics, Faculty of Physics},\orgname{ Alzahra University},\orgaddress{\state{Tehran},\country{Iran}}}


\abstract{Molecular absorption features constitute essential diagnostics of late-type stellar atmospheres, with titanium oxide (TiO) bands serving as sensitive tracers of effective temperature and magnetic activity. While near-infrared TiO indices are well established, visible-band diagnostics provide complementary constraints for stellar classification and parameter estimation. This study investigates the metallicity dependence of the B-index, a TiO absorption measure centered at 567 nm, through synthetic spectra generated with ATLAS9 and high-resolution observations from the HARPS spectrograph. The synthetic grid spans effective temperatures between $3500-4000 \ K$ and metallicities from [Fe/H] = -4 to +0.2, while the observational sample comprises 23 MK spectral types stars with effective temperatures between $3500-4000 \ K$. Both theoretical and empirical analyses demonstrate that the B-index exhibits negligible sensitivity to stellar metallicity, with only marginal correlations detected. Therefore, there is no requirement to include this feature in the calculation of the B-index.
The B-index's low sensitivity to metallicity reinforces its utility as an effective tool for spectral classification, stellar atmosphere modeling, and the study of magnetically active stars.}



\maketitle

\section{Introduction}\label{sec1}
Molecular absorption bands constitute one of the most distinctive spectroscopic features of late-type stellar atmospheres. Their presence reflects the relatively low photospheric temperatures that allow molecules to form and persist in equilibrium with atomic species. Among these, titanium oxide (TiO), vanadium oxide (VO), and cyanogen (CN) are particularly prominent, and their strengths provide valuable insight into the physical conditions of cool stars. These molecular signatures are also observed in magnetically active regions of earlier-type main-sequence stars (spectral types F-K), where localized cooling associated with starspots enhances their visibility  \citep{mirtorabi_2003}. Such molecular bands are therefore powerful probes of stellar photospheric conditions, as well as indicators of surface inhomogeneities linked to magnetic activity.

Titanium oxide is especially important in this context. The depth of its absorption bands has long been recognized as a diagnostic of fundamental stellar parameters, including effective temperature and surface inhomogeneities. In magnetically active stars, TiO absorption serves as a tracer of starspot coverage and variability, while in evolved pulsating giants, it has been used to study atmospheric structure and dynamical processes \citep{mirtorabi_2003}. TiO bands are also indispensable in spectral classification: their temperature sensitivity in the near-infrared provides the empirical foundation for the subdivision of M-type stars into finer spectral subgroups \citep{wing1992}. The bands first emerge at approximately spectral type K5, gradually intensify with decreasing temperature, and eventually dominate the near-infrared spectra of late M-type stars. Beyond spectral type M8, however, TiO features weaken, largely due to the increasing role of other molecular species and dust formation that obscure or blend the absorption \citep{rob1981}.

The diagnostic potential of TiO absorption motivated the development of specialized photometric techniques. In particular, \citet{wing1992} introduced a dedicated filter system optimized to isolate and measure TiO band strength in the near-infrared. This system has been widely employed in stellar classification and the study of cool stellar populations, as it provides a direct and temperature-sensitive measure of molecular absorption. Importantly, TiO absorption depth has been shown to be nearly insensitive to luminosity class, exhibiting comparable values in dwarfs, giants, and supergiants \citep{Wing1978}. This suggests that TiO absorption is governed primarily by temperature, with only secondary influence from surface gravity or global stellar structure, thus reinforcing its reliability as a temperature diagnostic across diverse stellar evolutionary stages.

With continued advances in spectroscopic instrumentation and theoretical models of line formation, attention has also turned to the characterization of TiO absorption in the visible spectrum. While near-infrared diagnostics remain fundamental, absorption features in the visible regime offer complementary avenues for spectral classification and parameter calibration, particularly with high-resolution spectrographs that routinely operate in this wavelength domain. Toward this goal, \citet{Bidaran_2016} introduced a photometric index centered at 567 nm, designed to quantify TiO absorption in the visible range. Using observations from the Kitt Peak National Observatory (KPNO) 2.1-m telescope \footnote{The KPNO 2.1-m telescope webpage: https://noirlab.edu/public/programs/kitt-peak-national-observatory/kpno-21m-telescope/}, they provided a preliminary calibration of this index against stellar temperature. However, earlier studies a drawing on the conclusions of \citet{Wing1978} atended to disregard the influence of metallicity, treating TiO absorption strength as essentially metallicity-independent.

This assumption, while convenient, merits closer scrutiny. Titanium oxide formation depends not only on the thermal conditions of the stellar photosphere but also on the availability of titanium and oxygen atoms, both of which are influenced by stellar metallicity. In metal-poor environments, the abundance of titanium may become a limiting factor, potentially altering the depth and detectability of TiO absorption bands. Conversely, in metal-rich stars, enhanced titanium abundance could increase band strength, complicating the interpretation of TiO indices as pure temperature diagnostics. Thus, the neglect of metallicity effects in earlier calibrations may introduce systematic uncertainties in the determination of stellar parameters, especially when applying TiO-based indices to large, chemically diverse stellar populations.

In light of these considerations, the present study revisits the diagnostic utility of TiO absorption in the visible spectrum with particular emphasis on metallicity dependence. We focus on the B-index, defined in the spectral region near 567 nm, as introduced by \citet{Bidaran_2016}. Our analysis employs both synthetic spectra generated from state-of-the-art stellar atmosphere models and an observational dataset of M-type stars obtained with the High Accuracy Radial velocity Planet Searcher (HARPS) spectrograph. The dual approach ensures a rigorous assessment of TiO absorption behavior across a wide range of effective temperatures and metallicities.

By explicitly quantifying the influence of metallicity on TiO absorption depth, we aim to refine the calibration of the B-index as a diagnostic tool for cool stellar atmospheres. This work thereby extends the foundational studies of \citet{Wing1978} and \citet{Bidaran_2016}, providing a more comprehensive framework for the application of TiO indices in stellar classification, parameter estimation, and the study of magnetically active and evolved cool stars. More broadly, such refinements contribute to the accurate characterization of late-type stellar populations, which are critical constituents of Galactic archaeology, exoplanet host surveys, and time-domain astrophysics.

\section{Titanium Oxide Index}\label{sec2}

TiO electronic transitions occur in both optical and near-infrared bands, dominated by allowed systems such as $\gamma (A^3 \Delta - X^3 \Delta)$, $\gamma' (B^3 \Pi - X^3 \Delta)$, and $\alpha (C^3 \Delta - X^3 \Delta)$, in which A, B and C (the three filters system of Wing) correspond to the excited levels with the same multiplicity as its ground state \citep{wing1992}. Forbidden transitions which are correspond to those that occur in transitions between the ground state and the non-equal multiplicity excited state are include mainly $\beta(c^1 \Phi - a^1 \Delta)$, $\delta(b^1 \Pi - a^1 \Delta)$ and $\phi(b^1 \Pi - d^1 \Sigma^+)$  \citep{Dobrodey2001}.

Among the available TiO line lists, \citet{Allard2000} compilation is considered the most accurate for atmospheric modeling, containing ~172 million lines. Synthetic spectra computed with ATLAS9 \citep{Kurucz1991,castelli2003} reveal that most TiO features in the optical are blended with other molecular absorptions. However, the TiO band centered at 567 nm remains relatively uncontaminated and serves as a suitable diagnostic. Vanadium oxide (VO) contributes negligible absorption in this region, with its strongest bands appearing only at longer wavelengths \citep{Kirkpatrick1991}.

Building on \citet{wing1992} photometric concept to measure the TiO strength in the near-infrared, \citet{Bidaran_2016} defined two new filters: D-band (559-575 nm) capturing TiO absorption and E-band (605-615 nm) sampling continuum. The B-index is then defined as the flux ratio:

\begin{eqnarray}
\label{eqn:1}
B-index=-2.5\log\frac{\int_{}^{}F_{D}(\lambda)~S_{D}(\lambda)~d\lambda}{\int_{}^{}~F_{E}(\lambda)~S_{E}(\lambda)~d\lambda}
\end{eqnarray}

where $F_{D}(\lambda)$ and $F_{E}(\lambda)$ present the flux collected in each of the filters and $S_{D}(\lambda)$ 
and $S_{E}(\lambda)$ are the appropriate filter response functions in each filter.

\begin{figure}
\centering
\includegraphics[width=1\textwidth]{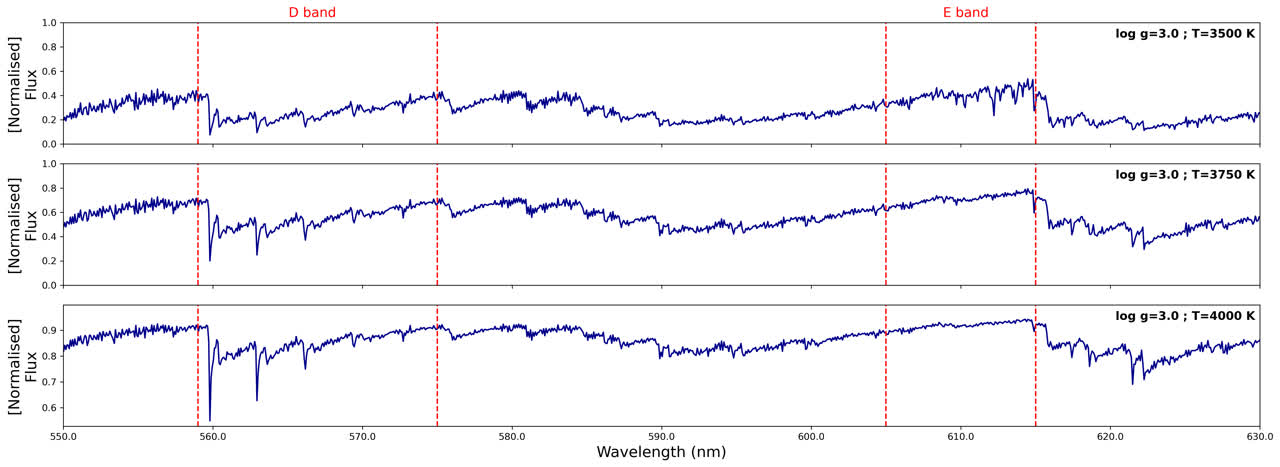}
\caption{The synthetic spectra of visual absorption features of TiO$\lambda567~nm$ molecule in the atmosphere of cool stars, with different temperature and same gravity and metallicity. All the data has been produced using ATLAS 9 software.
As it is obvious from the figure, TiO absorption bands deepens as temperature decreases declares that TiOs population increases as the star becomes cooler. The selected band passes D and E are also shown in the panels. In all three panels flux is normalized to continume. Filter D measures 559-575$ \mathrm{nm}$.filter E measures 605-615 $\mathrm{nm}$. D and E depict regions that were used to measure B-index.}
\label{FIG:1}
\end{figure}

Synthetic spectra show that the B-index decreases monotonically with increasing effective temperature from 3500-4000 $K$, indicating its robustness as a temperature-sensitive diagnostic (see Figure~\ref{FIG:1}). While minor atomic and molecular contaminants exist, the smooth index temperature trend demonstrates its reliability \citep{Bidaran_2016}. Thus, the B-index provides a practical visual diagnostic for TiO absorption in cool stars.

\section{Data, Calculations and discussion}\label{sec3}
\subsection{Synthetic data}\label{subsec1}
We computed a comprehensive grid of stellar atmosphere models for cool dwarf stars using the ATLAS9 software package \citep{Kurucz1991,castelli2003}. 

In the 3500 - 4000 K regime, where TiO, VO, and $H_{2}O$ opacities become strong and dust begins to appear in dwarfs, ATLAS9 is selected over MARCS or PHOENIX despite their superior molecular linelists, because our analysis deliberately avoids saturated bandheads, relies on empirically calibrated and updated ODFs, and requires a homogeneous treatment across a broad temperature range (up to 6000 K) to prevent systematic offsets. Moreover, ATLAS9's computational efficiency, reducing weeks of CPU time to hours, enables robust statistical error estimation for our large sample. While acknowledging that BT-Settl/PHOENIX are essential for brown dwarfs and MARCS for metal-poor or C-rich stars, we conclude that ATLAS9 is the optimal pragmatic choice for our solar-metallicity, high-gravity dwarf-star analysis, offering the best trade-off between accuracy, grid consistency, and numerical feasibility \citep{boz2015}.

The models were designed to explore the parameter space where titanium oxide (TiO) absorption plays a dominant role in shaping the emergent stellar spectra. The grid spans metallicities from $\left[\mathrm{Fe}/\mathrm{H}\right] = -4.0$ to $+0.2$ including -4.0, -2, -1, -0.5, 0, 0.2, 0.5, thereby encompassing extremely metal-poor stars up to slightly supersolar compositions. Effective temperatures of $T_{\mathrm{eff}} = 3500,\  3750$ and $4000 ,\mathrm{K}$ were selected, since these represent the range where TiO molecular bands become particularly prominent in late-type dwarfs.

\begin{figure}
\centering
\includegraphics[width=0.9\textwidth]{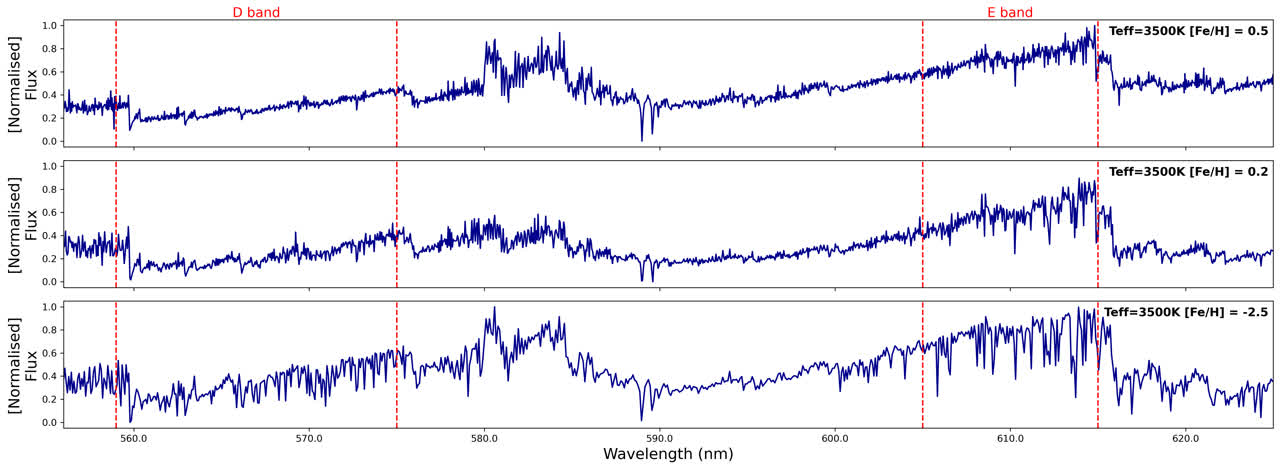}
\caption{The synthetic spectra of visual absorption features of TiO$\lambda567~nm$ molecule in the atmosphere of cool stars, with different metallicity and same temperature. All the data has been produced using ATLAS 9 software. The selected band passes D and E are also shown in the panels. Filter D measures 559-575 $\mathrm{nm}$.filter E measures 605-615 $\mathrm{nm}$.D and E depict regions that were used to measure B-index.}
\label{FIG:2}
\end{figure}

From this grid, a total of 23 high-resolution (R= 600,000) synthetic spectra were generated, covering the optical range where TiO absorption is most significant. Figure~\ref{FIG:2} illustrates three representative synthetic spectra computed at $T_{\mathrm{eff}} = 3500 \ \mathrm{K}$ for different metallicities. The comparison clearly reveals the influence of chemical composition on the overall strength and morphology of molecular absorption bands. However, it is noteworthy that the absorption feature centered near $567 \mathrm{nm}$ remains essentially unchanged across all metallicities considered at this temperature.

\begin{figure}
\centering
\includegraphics[width=0.9\textwidth]{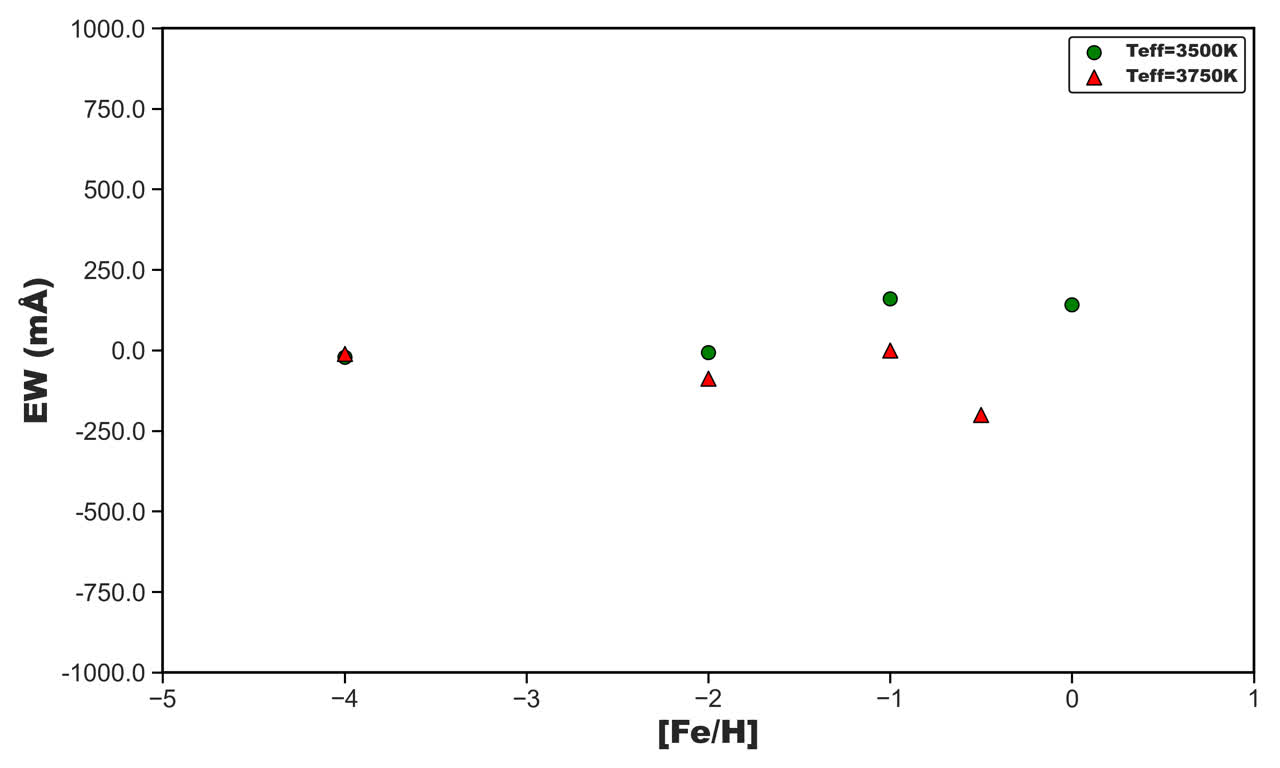}
\caption{Equivalent width of the TiO$\lambda567~nm$ absorption line versus metallicity at temperatures of 3500 and 3750 K.}
\label{FIG:3}
\end{figure}

For the quantitative analysis, the equivalent width of the absorption line (TiO$\lambda567~nm$) was measured. Each panel in Figure~\ref{FIG:3} shows the variation of the absorption line (TiO$\lambda567~nm$) equivalent width at a specific temperature across different metallicities. As can be seen from each panel of Figure~\ref{FIG:3}, the equivalent width of the absorption line does not change significantly with varying metallicities at a given temperature.

This behavior suggests that the formation of the $567 \mathrm{nm}$ line is relatively insensitive to variations in bulk metallicity, in contrast to the observable dependence of TiO absorption to effective temperature specifically in near infrared \citep{mirtorabi_2003}, \citep{wing1992}. The relative stability of this feature implies that it may serve as a robust diagnostic of photospheric conditions, particularly effective temperature, largely independent of the stars chemical composition. Such metallicity-independent diagnostics are essential for constraining stellar parameters in metal-poor populations where spectroscopic analyses are otherwise complicated by weakened molecular absorption.

\subsection{Observational data}\label{subsec2}

To extend the analysis to a more experimental ground, we focus on a stellar sample that includes 23 stars observed with the High Accuracy Radial velocity Planet Searcher (HARPS \footnote{The HARPS webpage: http://www.eso.org./instruments/harps/}) instrument (see the first column of Table~\ref{tbl:01}). HARPS is a high-resolution, fiber-fed, cross-dispersed echelle spectrograph mounted on the 3.6-meter Cassegrain telescope at the European Southern Observatory (ESO) in La Silla, Chile. Since its commissioning in 2003, it has become one of the most advanced and reliable facilities for precision spectroscopy.
HARPS was originally designed for the detection and characterization of exoplanets through the radial-velocity method \citep{mayor2003}. This level of precision allows the detection of Earth-mass exoplanets orbiting solar-type stars, but it also provides unique opportunities for a wide range of stellar astrophysics applications. The instrument covers a spectral range of $378$-$691\mathrm{nm}$, which includes numerous atomic and molecular absorption features essential for analyzing the physical and chemical properties of late-type stars. With a resolving power of $R \approx 115{,}000$, HARPS is able to resolve fine details in line profiles, enabling accurate determinations of stellar abundances, temperature-sensitive indices, and activity diagnostics.
The stellar sample analyzed here was selected according to several criteria. Only stars with high signal-to-noise ratio $(S/N > 100)$ spectra were included to ensure robust parameter determination. Preference was given to late-type dwarfs spanning a wide range of metallicities with $logg=0.5 - 4$, as these stars exhibit strong molecular absorption features, particularly TiO bands, which serve as powerful diagnostics of effective temperature and photospheric structure. Moreover, sampling stars across different metallicities allows us to investigate the influence of chemical composition on molecular band strengths, which is essential for both stellar parameter calibration and stellar population studies. Stars with strong variability or poorly constrained photometry were excluded to minimize systematic uncertainties.

The inclusion of HARPS data under these carefully chosen criteria provides a highly reliable dataset, enabling precise investigations of the photospheric and chromospheric properties of the selected targets.
The spectra are accessible from the ESO archives.

\subsection{Calculation of B-index}\label{subsubsec3}
\begin{figure}
\centering
\includegraphics[width=0.9\textwidth]{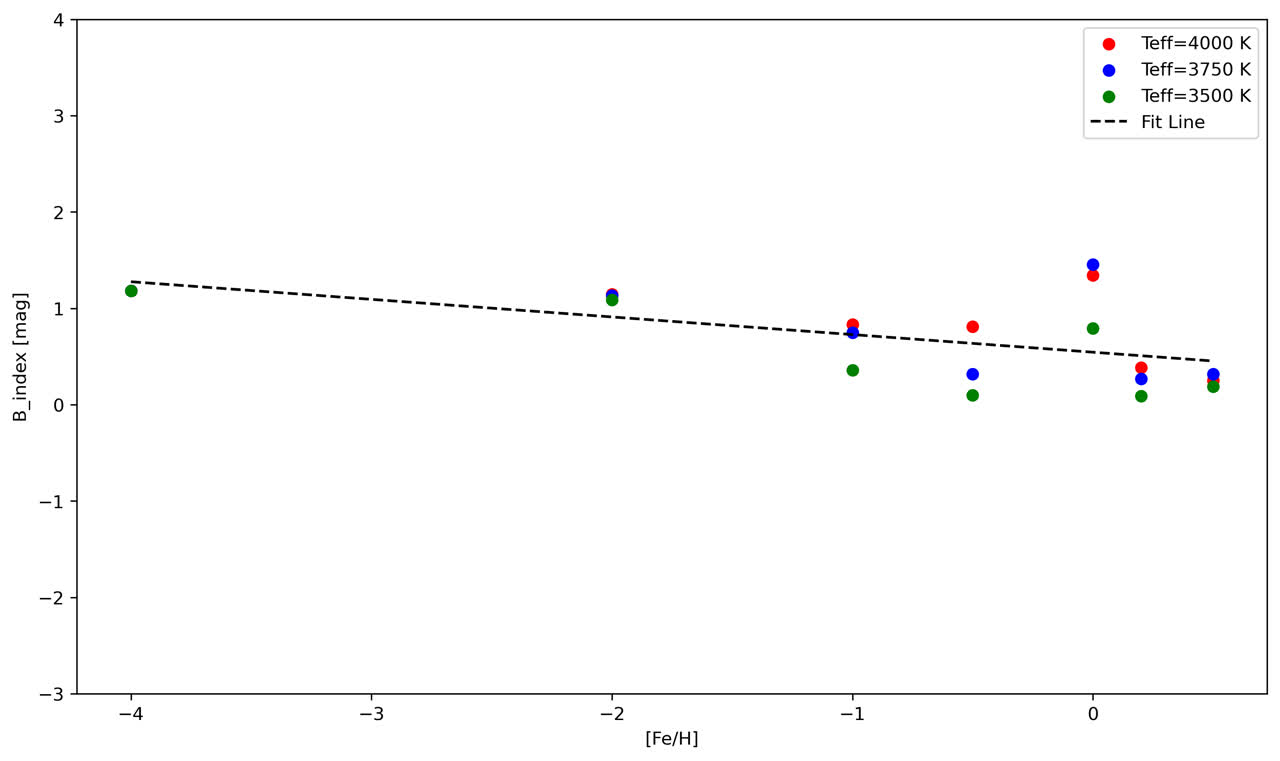}
\caption{B-index plot relative to metallicity for synthetic spectrum produced from ATLAS9.}
\label{FIG:4}
\end{figure}

To assess the impact of metallicity on the B-index, we first computed this index for the entire grid of synthetic spectra using Equation~\ref{eqn:1}. The results are illustrated in figure~\ref{FIG:4}, which displays the variation of the B-index as a function of metallicity for the three effective temperatures considered. As is evident from the figure, the dependence of the B-index on metallicity is extremely weak across the full temperature range. A linear least-squares fit to the data yields the relation

\begin{eqnarray}
\label{eq2}
B-index = -0.183 [\frac{Fe}{H}] +0.544
\end{eqnarray}

\noindent indicating only a marginal sensitivity to chemical composition in the synthetic models. This finding suggests that the B-index is primarily governed by photospheric temperature rather than by metallicity effects.

In addition to the synthetic dataset, we measured the B-index for the observed stellar sample. For each of the 23 HARPS stars, the depth of the titanium oxide absorption feature at 567 nm was determined using Equation~\ref{eqn:1}. These values are reported in the fourth column of Table~\ref{tbl:01}. For reference, the second column of the table lists the spectral classifications of the stars, as determined by \citet{Sousa_2008} using the same HARPS spectra, while the third column provides their metallicities.

\begin{figure}
\centering
\includegraphics[width=0.9\textwidth]{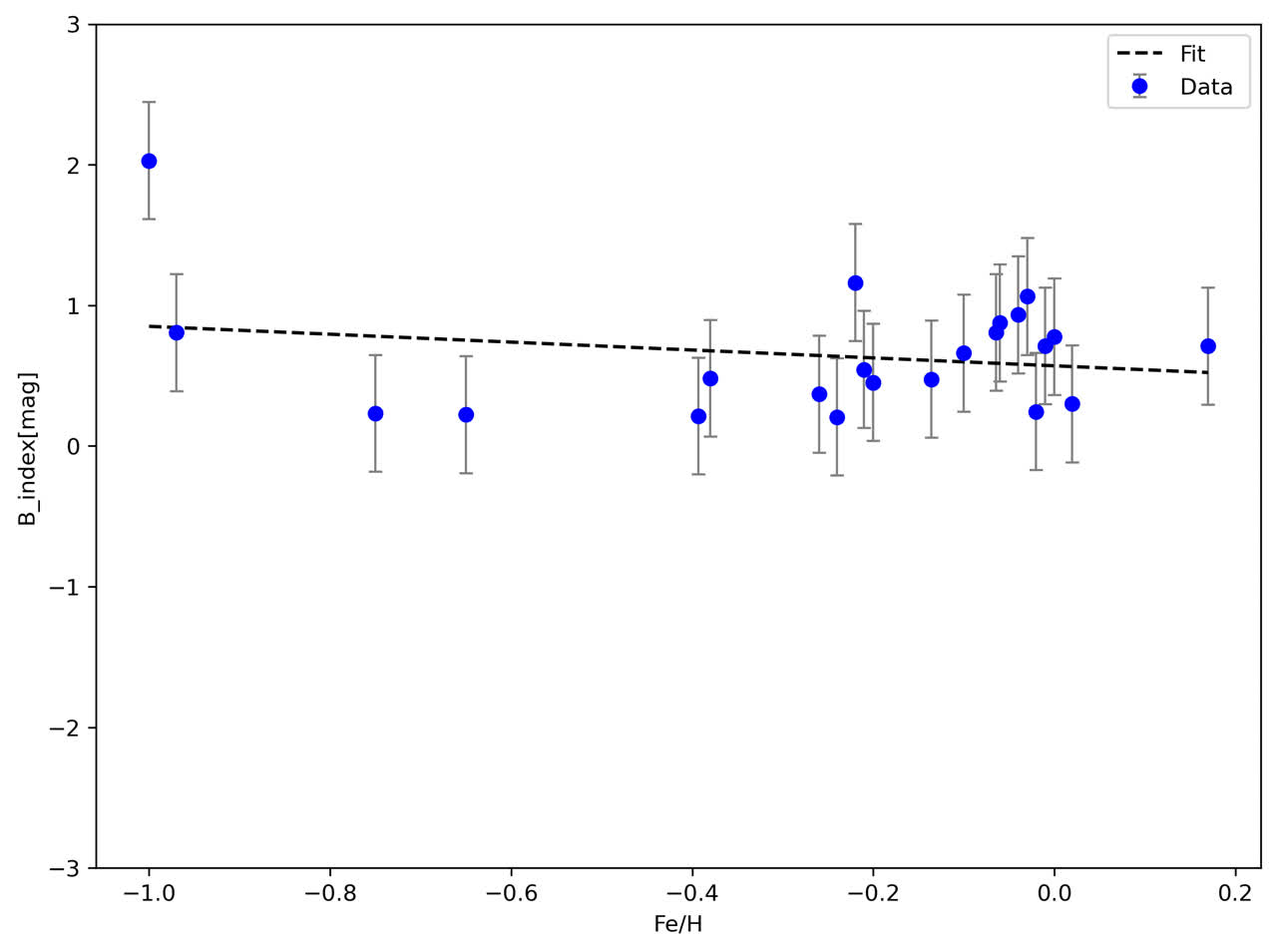}
\caption{B-index plot relative to metallicity for the HARPS sample stars.}
\label{FIG:5}
\end{figure}

Figure~\ref{FIG:5} presents the observed B-index as a function of stellar metallicity for the HARPS sample. Consistent with the synthetic analysis, the empirical data demonstrate that metallicity has a negligible influence on the B-index. The best-fitting linear relation with R-squared of 0.416 is found to be

\begin{eqnarray}
\label{eq3}
B-index= -0.281 [\frac{Fe}{H}] +0.570,
\end{eqnarray}

Pearson correlation coefficient is -0.211 \noindent which again confirms the weak sensitivity of the index from chemical composition. The slope of the observed relation is even smaller than that obtained from the synthetic models, reinforcing the robustness of this diagnostic against metallicity variations.

\section{Summary and Discussion}\label{sec3}
Molecular absorption features in late-type stellar atmospheres serve as powerful diagnostics of stellar parameters, particularly effective temperature and surface inhomogeneities. Titanium oxide (TiO) is among the most significant molecular species, its absorption bands providing valuable constraints on the physical conditions of cool stars. Traditionally, TiO indices have been exploited in the near-infrared domain, where their sensitivity to temperature enables robust spectral classification and stellar parameterization. More recently, attention has shifted to the visible regime, where a photometric index, the B-index, centered near 567 nm, has been proposed as a temperature diagnostic. However, a key uncertainty concerns the degree to which this index is affected by stellar metallicity, since molecular band formation inherently depends on the abundances of constituent elements.

The present study addresses this question through a twofold methodology, combining synthetic stellar spectra generated with the ATLAS9 code and high-resolution observational data obtained with the HARPS spectrograph. The synthetic grid encompasses metallicities from [M/H] = -4.0 to +0.2 and effective temperatures between $3500 - 4000 \ K$, corresponding to the range where TiO absorption is most prominent in late type dwarfs. Parallelly, a carefully selected sample of 23 M-K type stars with high signal to noise HARPS spectra was analyzed to empirically evaluate the B-index across a range of metallicities.

The theoretical spectra reveal that while many TiO features vary appreciably with metallicity, the absorption band at 567 nm remains remarkably stable across the entire metallicity grid. This insensitivity was quantified by fitting the B-index as a function of [Fe/H], yielding only a marginal dependence (equation \ref{eq2}). The observational dataset corroborates these results, with HARPS measurements producing an even weaker metallicity dependence (equation \ref{eq3}). Thus, both synthetic modeling and empirical analysis converge on the conclusion that B-index's dependence on metallicity is weak.
To elaborate, we do not claim that B-index is independent of metallicity; rather, our initial intention was to incorporate a sensitivity factor into B-index, though we subsequently found this effect to be weak.
This finding has significant implications. First, it strengthens the status of the B-index as a reliable diagnostic of effective temperature, largely immune to chemical composition. This robustness simplifies its application to chemically diverse stellar populations, including metal-poor environments where metallicity-sensitive indices often fail. Secondly, the insensitivity to metallicity enhances the utility of the B-index in large-scale stellar surveys, where homogeneity of diagnostics across broad metallicity ranges is essential for unbiased population studies. Moreover, given that TiO absorption is also a sensitive tracer of magnetic activity and starspot coverage, low sensitivity to metallicity ensures that variability analyses based on the B-index will not be confounded by underlying chemical differences.

Nevertheless, caution is warranted. While the results convincingly demonstrate the low sensitivity to metallicity at 567 nm, TiO formation is still fundamentally tied to titanium and oxygen abundances. Thus, at extreme metallicity regimes, or in chemically peculiar stars, secondary effects may still emerge. Additionally, blending from other molecular and atomic species, though minimal at this wavelength, could introduce subtle systematic errors in cases of low-resolution or low signal-to-noise data. Future work may benefit from extending the calibration to cooler stars approaching the brown dwarf regime, where dust formation further complicates TiO visibility.

 \section*{Acknowledgments}
Based on observations collected with the {\it HARPS \/}  spectrograph on the 3.6-m telescope at {\it La Silla \/} Observatory,  European Southern Observatory, Chile, Science Archive Facility.

\section*{Data Availability}

The data underlying this article are available in  the ESO Science Archive Facility in "https://archive.eso.org/scienceportal/home".









\begin{table}[h]
\centering
\caption{The Results.} \label{tbl:01}
\begin{tabular}{@{}lllllllll@{}}
\hline
$Stars$ & $Spectral  $  &$[Fe/H]$&$B-Index$& $Ref, [Fe/H]$\footnotemark[3]   \\ 
 $ name$  & $type\footnotemark[2]  $&     &  [mag]   &   &    &  &  \\
\hline
HD29139 	&K5 & -0.2&0.4520&  \citep{ros2024}   \\
HD99167 & K4 &  -0.38&	0.4813	&\citep{mc1990} \\
HD148513&K3& -0.02  &0.2447&  \citep{sou2022}   \\
HD167006 &	M3 &    -0.97&	  0.8054    &\citep{ga2021}   \\
HD180928 &K4&   -0.75& 0.2322&  \citep{ga2021}    \\
HD177940 &M7 &-1&  2.0292&   \citep{sou2022}   \\
HD189319& M0   &-0.26&  0.3683&   \citep{sou2024}  \\
HD225212&K3& 0.02& 0.2988&  \citep{sou2022}  \\
HD146051&M0.5&-0.136& 0.4751&   \citep{and2016}   \\
HD39801 &	 M1-M2   &-0.064&0.8077&   \citep{tan2025}  \\
HD99998 &	 K3 &-0.393 &	0.2121&  \citep{sou2024}  \\
HD118100 &	K5 &-0.65&0.2230	&  \citep{ga2021}  \\
HD123657 &	 M4.5  & -0.03 & 1.0631&  \citep{cha2020}   \\
HD123934 &	M2   & -0.21  & 0.544&   \citep{and2016}   \\
HD131918 & K4 &-0.24 &0.2056& \citep{ta2023}  \\
HD42543 &	M0 &	0.17&0.7095&\citep{pr2011}  \\
HD200527 &	 M4.5 &-0.04   &0.9327& \citep{sm1986}  \\
HD204445&	M1 &	 -0.01  &0.7110& \citep{yu2018}   \\
HD221615&	M5   & -0.22 &	1.1615& \citep{yu2018}  \\
HD172816&	 M4  & 0.71 & 1.0358&   \citep{ga2021} \\
HD112142&	M3   &0&0.7764&  \citep{cha2020} \\
HD196777 &	 M1  	& -0.1   &  0.6598&  \citep{adel1973}    \\
HD112300	&M3&   -0.06  &0.8754& \citep{sousa2024}  \\
\hline
\end{tabular}


\end{table}

\footnotetext[2]{the spectral classifications of the stars, as determined by \citet{Sousa_2008} using the same HARPS spectra}
\footnotetext[3]{References for metallicity}


\end{document}